\documentclass[aps,prb,preprint,groupedaddress]{revtex4}
\usepackage{graphicx}
\usepackage{amsmath}
\usepackage{wrapfig}
\usepackage{multirow}

\begin{document}

% BOLD SYMBOLS
\newcommand{\GGG}{Gd$_3$Ga$_5$O$_{12}$}

\newcommand{\AAA}{\mathbf{A}}
\newcommand{\AAn}{\mathbf{A}^\notop}
\newcommand{\aaa}{\mathbf{a}}
\newcommand{\aan}{\mathbf{a}^\notop}
\newcommand{\BBB}{\mathbf{B}}
\newcommand{\BBn}{\mathbf{B}^\notop}
\newcommand{\bbb}{\mathbf{b}}
\newcommand{\bbn}{\mathbf{b}^\notop}
\newcommand{\ccc}{\mathbf{c}}
\newcommand{\CCC}{\mathbf{C}}
\newcommand{\CCn}{\mathbf{C}^\notop}
\newcommand{\DDD}{\mathbf{D}}
\newcommand{\DDn}{\mathbf{D^\notop}}
\newcommand{\ddd}{\mathbf{d}}
\newcommand{\ddn}{\mathbf{d^\notop}}
\newcommand{\EEE}{\mathbf{E}}
\newcommand{\EEn}{\mathbf{E^\notop}}
\newcommand{\eee}{\mathbf{e}}
\newcommand{\een}{\mathbf{e}^\notop}
\newcommand{\FFF}{\mathbf{F}}
\newcommand{\FFn}{\mathbf{F}^\notop}
\newcommand{\FFFe}{\mathbf{F}^\notop_\textrm{el}}
\newcommand{\fff}{\mathbf{f}}
\newcommand{\ffn}{\mathbf{f}^\notop}
\newcommand{\fffe}{\mathbf{f}^\notop_\textrm{el}}
\newcommand{\GGn}{\mathbf{G}}
\newcommand{\ggn}{\mathbf{g}^\notop}
\newcommand{\hhh}{\mathbf{h}}
\newcommand{\HHh}{\mathbf{H}}
\newcommand{\JJJ}{\mathbf{J}}
\newcommand{\JJn}{\mathbf{J}^\notop}
\newcommand{\JJJe}{\mathbf{J}^\notop_\textrm{el}}
\newcommand{\kkk}{\mathbf{k}}
\newcommand{\kkn}{\mathbf{k}^\notop}
\newcommand{\KKK}{\mathbf{K}}
\newcommand{\KKn}{\mathbf{K}^\notop}
\newcommand{\mmm}{\mathbf{m}}
\newcommand{\MMM}{\mathbf{M}}
\newcommand{\MMn}{\mathbf{M}^\notop}
\newcommand{\nnn}{\mathbf{n}}
\newcommand{\PPP}{\mathbf{P}}
\newcommand{\PPn}{\mathbf{P}^\notop}
\newcommand{\ppp}{\mathbf{p}}
\newcommand{\ppn}{\mathbf{p}^\notop}
\newcommand{\QQQ}{\mathbf{Q}}
\newcommand{\QQn}{\mathbf{Q}^\notop}
\newcommand{\qqq}{\mathbf{q}}
\newcommand{\qqn}{\qqq^\notop}
\newcommand{\RRR}{\mathbf{R}}
\newcommand{\RRn}{\RRR^\notop}
\newcommand{\rrr}{\mathbf{r}}
\newcommand{\rrn}{\rrr^\notop}
\newcommand{\sss}{\mathbf{s}}
\newcommand{\SSS}{\mathbf{S}}
\newcommand{\ssn}{\sss^\notop}
\newcommand{\uuu}{\mathbf{u}}
\newcommand{\uun}{\mathbf{u}^\notop}
\newcommand{\unn}{{u}^\notop}
\newcommand{\vvv}{\mathbf{v}}
\newcommand{\vvn}{\mathbf{v}^\notop}
\newcommand{\vnn}{{v}^\notop}
\newcommand{\www}{\mathbf{w}}
\newcommand{\wwn}{\mathbf{w}^\notop}
\newcommand{\wnn}{{w}^\notop}

\newcommand{\eqlab}[1]{\label{eq:#1}}
\renewcommand{\eqref}[1]{(Equation~\ref{eq:#1})}
\newcommand{\eqsref}[2]{(Equations~(\ref{eq:#1}) and~(\ref{eq:#2}))}
\newcommand{\eqsrefs}[3]{(Equations~(\ref{eq:#1}),~(\ref{eq:#2})  and~(\ref{eq:#3}))}
\newcommand{\exref}[1]{\ref{ex:#1}}
\newcommand{\exlab}[1]{\label{ex:#1}}
\newcommand{\figref}[1]{Fig.~\ref{fig:#1}}
\newcommand{\figsref}[2]{Figs.~\ref{fig:#1} and~\ref{fig:#2}~}
\newcommand{\figlab}[1]{\label{fig:#1}}
\newcommand{\chapref}[1]{Chapter~\ref{chap:#1}}
\newcommand{\chapsref}[2]{Chapters~\ref{chap:#1} and~\ref{chap:#2}}
\newcommand{\chaplab}[1]{\label{chap:#1}}
\newcommand{\secref}[1]{Section~\ref{sec:#1}}
\newcommand{\secsref}[2]{Sections~\ref{sec:#1} and~\ref{sec:#2}}
\newcommand{\seclab}[1]{\label{sec:#1}}
\newcommand{\tabref}[1]{Table~\ref{tab:#1}}
\newcommand{\tabsref}[2]{Tables~\ref{tab:#1} and~\ref{tab:#2}}
\newcommand{\tablab}[1]{\label{tab:#1}}

\title{Updating the phase diagram of the archetypal frustrated magnet Gd$_3$Ga$_5$O$_{12}$ }
\author{P. P. Deen}
\affiliation{ESS, Tunav\"{a}gen 24, 223 63 Lund, Sweden}
\affiliation{Nanoscience Center, Niels Bohr Institute, University of Copenhagen, DK-2100 Copenhagen {\O}, Denmark}
\author{O. Florea}
\affiliation{Institut N{\'e}el, CNRS $\&$ Universit{\'e}, Joseph Fourier, BP 166, 38042 Grenoble Cedex 9, France}
\author{E. Lhotel}
\affiliation{Institut N{\'e}el, CNRS $\&$ Universit{\'e}, Joseph Fourier, BP 166, 38042 Grenoble Cedex 9, France}
\author{H. Jacobsen}
\affiliation{Nanoscience Center, Niels Bohr Institute, University of Copenhagen, DK-2100 Copenhagen {\O}, Denmark}
\affiliation{ESS, Tunav\"{a}gen 24, 223 63 Lund, Sweden}

%\author{H. Mutka}
%\affiliation{Institut Laue-Langevin, 6 rue Jules Horowitz, 38042 Grenoble, France}
%\author{O. A. Petrenko}
%\affiliation{Department of Physics, University of Warwick, Coventry, CV4 7AL, United Kingdom}

\date{\today}

\begin{abstract}

The applied magnetic field and temperature phase diagram of the archetypal frustrated magnet, Gd$_{3}$Ga$_{5}$O$_{12}$, has been reinvestigated using single crystal magnetometry and polarised neutron diffraction. 
The updated phase diagram is substantially more complicated than previously reported and can be understood in terms of competing interactions with loops of spins, trimers and decagons, in addition to competition and interplay between antiferromagnetic, incommensurate and ferromagnetic order. Several additional distinct phase boundaries are presented. The phase diagram centers around a multiphase convergence to a single point at 0.9 T and $\sim$ 0.35 K, below which, in temperature, a very narrow magnetically disordered region exists. These data illustrate the richness and diversity that arises from frustrated exchange on the 3 dimensional hyperkagome lattice.

\end{abstract}
\pacs{75.30.Kz, 75.30.Cr, 61.04.fm}
%Magnetic phase transitions, 75.30.Kz
%magnetic susceptibility, 75.30.Cr
% neutron diffraction, 61.05.fm
%quantum spin frustration, 75.10.Jm
\maketitle
\section{Introduction}

Frustration is ubiquitous in nature and drives the physical behaviour in materials ranging from liquid crystals and polymers to compounds with
localised magnetic moments \cite{Lotz2012, Araki2011, Balents2010, Gardner2010}. The study of magnetic frustration, in
particular, is proving very fruitful in the development of a more general understanding of frustrated phenomena. Magnetic frustration is governed by
the connectivity of the many degenerate spin configurations in its ground
state manifold with geometric frustration borne out of the  crystal structure topology.  The multiplicity of ground states at the lowest temperatures, $T$ $\rightarrow$ 0~K,  can lead to persistent dynamic magnetic spins \cite{Lacroix}, a most exotic state.\\

The rare earth garnet Gd$_{3}$Ga$_{5}$O$_{12}$ (GGG) is unique since it offers a rare opportunity to study frustration on a double hyperkagome structure, a 3D kagome lattice of interconnected triangles.  GGG is regarded as the archetypal frustrated compound since, unlike its many counterparts, it does 
% need to check the compounds, and add ref. 
 not revert to an ordered state via an "order by disorder" transition \cite{Villain1980}. Indeed there is no sign of long range order in GGG down to 0.025~K despite
a Curie-Weiss temperature of $\theta_{CW}$ $\sim$ -2~K \cite{Onn1966, Kinney1979} indicating strongly frustrated antiferromagnetic (AF) correlations.  In contrast, the rare earth counterparts Dy$_3$Ga$_5$O$_{12}$ and Ho$_3$Ga$_5$O$_{12}$ order at relatively high temperatures thereby highlighting the unique nature of the magnetic exchange interactions in GGG \cite{Onn1966}. \\

The leading magnetic interactions in GGG are the  near neighbor, $J_{1}$ = -0.107~K \cite{Hov1980,Kinney1979}, and the dipole exchange interactions, $D$ = 0.0457 K \cite{PetrenkoPRB_2000}.  Adjacent triangle and sublattice exchange interactions, $J_{2}$ and $J_{3}$ respectively, are an order of magnitude smaller, $J_{2}$ $\sim$ -0.005 and $J_{3}$ $\sim$ 0.010K \cite{YavorskiiPRL2006}.  In GGG the magnetic Gd$^{3+}$ spins ($S$ = 7/2) are considered as Heisenberg spins with single ion anisotropy of less than 0.04 K. However the non-negligible dipole exchange and the local crystal field environment could lead to anisotropy \cite{Overmeyer}.  The inherent spin Hamiltonian of the localised Gd$^{3+}$ moment on the hyperkagome lattice gives rise to magnetic short range order (SRO) below $T$ $\sim$ 3~K \cite{Kinney1979, Schiffer1995, Petrenko1998, PetrenkoPRB_2000}.  Below about 0.018~K, a spin glass phase has been observed by ac susceptibility and specific heat at ambient fields \cite{Schiffer1995} and verified by Petrenko {\it et al.} \cite{Petrenko1998} using neutron scattering.  Neutron scattering revealed sharper but not resolution limited magnetic diffraction peaks in addition to SRO for T~$<$~0.14 K \cite{Petrenko1998}. These correlations are understood to result from the long range nature of the dipole exchange interactions perturbed by the very weak exchange interactions, J$_{2}$ and J$_{3}$\cite{YavorskiiPRL2006}, and may give rise to a quantum protectorate of ten ion spin clusters \cite{Ghosh2008}.

The first (H, T)  phase diagram of GGG indicated a AF dome with strong anisotropy such that applying the field along the [1 0 0] creates an  AF dome for $T$ $<$ 0.4  and 0.85 $<$ $\mu_{0}$H $<$ 1.8~T while applying the field along [1 1 1] results in an AF  dome for $T$ $<$ 0.3~K and  0.9 $<$ $\mu_{0}$H $<$ 1.4~T \cite{Hov1980}. The anisotropy of the phase diagram was further probed via magnetic susceptibility, specific heat and  single crystal neutron scattering \cite{PetrenkoHFM, Schiffer1994} and revealed that the phase diagram is much more complex than previously reported. The boundary between the disordered region above the spin glass phase and the long ranged AF dome showed considerable similarities with the melting curve of $^{4}$He indicating unusual magnetic behaviour in this region \cite{Tsui1999}. In contrast, muon spin relaxation and M\"{o}ssbauer spectroscopy indicate a temperature spin relaxation  indicative of slow spin fluctuations down to 0.025~K and up to 1.8~T \cite{Dunsiger2000, Marshall2002, Bonville}. The contradiction between a long range ordered and dynamic magnetic state can be understood in terms of the time scales probed by the various techniques such that the system appears paramagnetic on the fast timescale of the muon but looks static within the energy resolution of the neutron. This has been observed in the frustrated magnet Tb$_{2}$Ti$_{2}$O$_{7}$ for which it was shown that there were many length and timescales to consider \cite{TTO2012}.

%The phase diagram was further extended using susceptibility and specific heat measurements by Schiffer {\it et al.} \cite{Schiffer1995} to include the spin glass phase, T $<$ 0.14~K,  at ambient fields as verified  by Petrenko {\it et al.} \cite{Petrenko1998}  using neutron scattering. 
This study revisits the temperature and field dependent behaviour of GGG  through macroscopic magnetic measurements of both powdered and single crystal samples and a polarised neutron scattering study of a powdered sample. In this study the field is applied along the [1 1 0] crystallographic direction to complement previous results. The phase diagram presents several additional phases for this direction of the applied magnetic field. These phases are also observed for the powder  sample, but are broadened suggesting different characteristic field dependencies as a function of direction, as reported previously. Neutron powder diffraction correlate closely with the macroscopic measurements. 

\section{Magnetization measurements}
\label{Mag}
\subsection{Experimental details}
Magnetization measurements were performed by the extraction method down to 0.08 K and up to 8 T, using a superconducting quantum interference device magnetometer equipped with a miniature dilution refrigerator, developed at the Institut N\'eel \cite{Paulsen}. Two \GGG samples have been measured: i) a 8.52~mg single crystal cut from a commercial $(10~\times~10~\times~0.4)$~mm$^{3}$ substrate (Impex High Tech), measured along the [1 1 0] direction and ii) a 17.13~mg $^{160}$Gd isotope powdered sample. Neutron scattering experiments were performed on a powder sample of the same batch for a direct comparison. Magnetization measurements performed on the powder are in quantitative agreement with single crystal measurements. However, the features are broadened, certainly due to the distribution of crystallites and the presence of a small anisotropy ($\sim$ 0.04 K \cite{Overmeyer}).
%It is rather surprising that is possible to directly compare the powder and single crystal magnetisation measurements considering the large variation in the precise details of the phase diagrams with the application of magnetic fields along different crystallographic directions \cite{Hov, PetrenkoHFM}.
%We decided not to put this!
The results presented below are from the single crystal sample. The field was applied in the plane of the substrate, so demagnetization corrections are negligible.

\subsection{Phase diagram via magnetization measurements}

\begin{figure}[h]
\includegraphics[keepaspectratio=true, width=8cm]{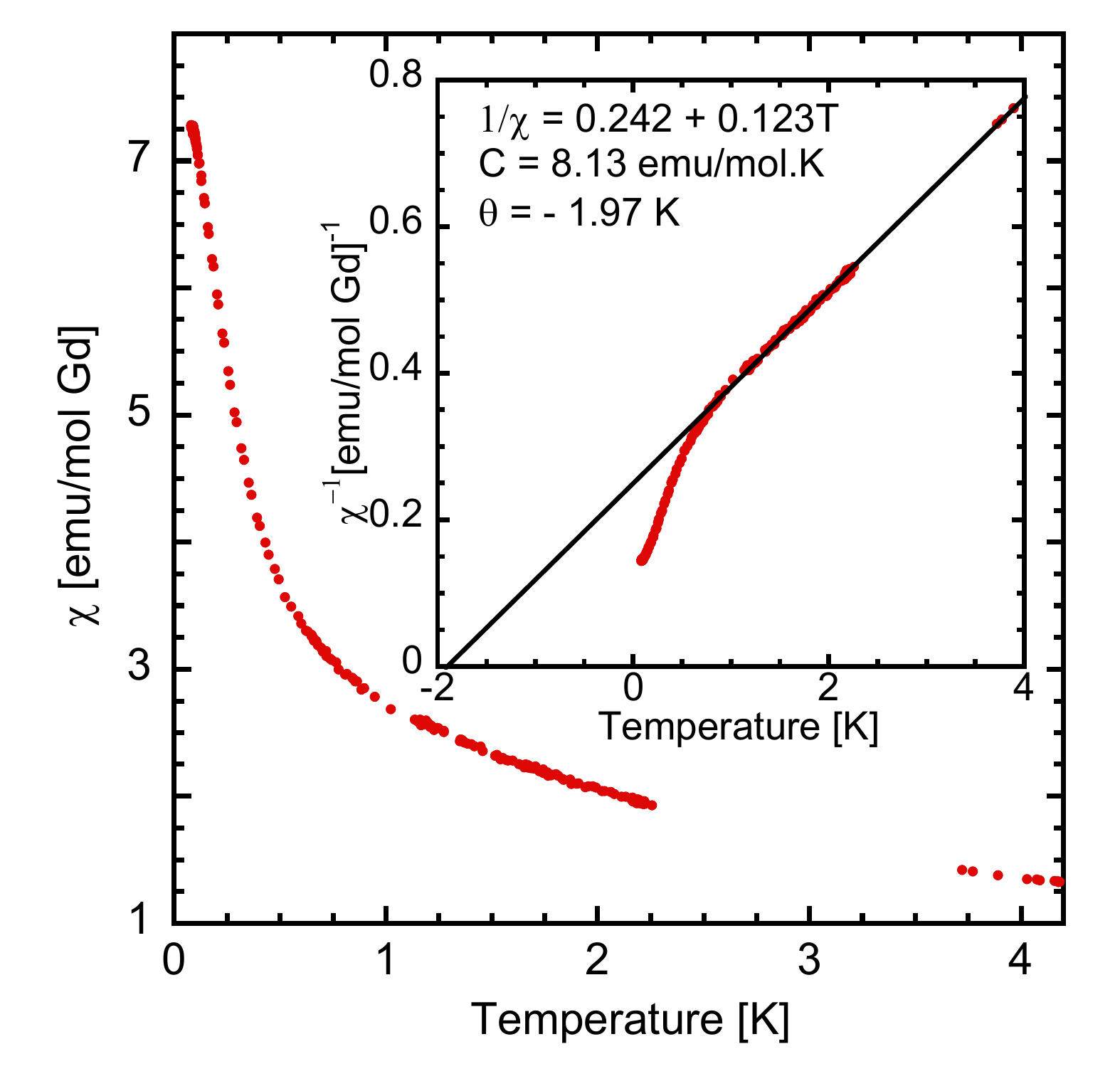}
\caption{$\chi$ vs $T$ at 0.01 T with H $\parallel$ to [1 1 0] for 0.08~$< T <$~4.2 K. Inset: $\chi^{-1}$ vs. $T$.}
\label{figXT}
\end{figure}

\begin{figure}[h]
\includegraphics[keepaspectratio=true, width=8cm]{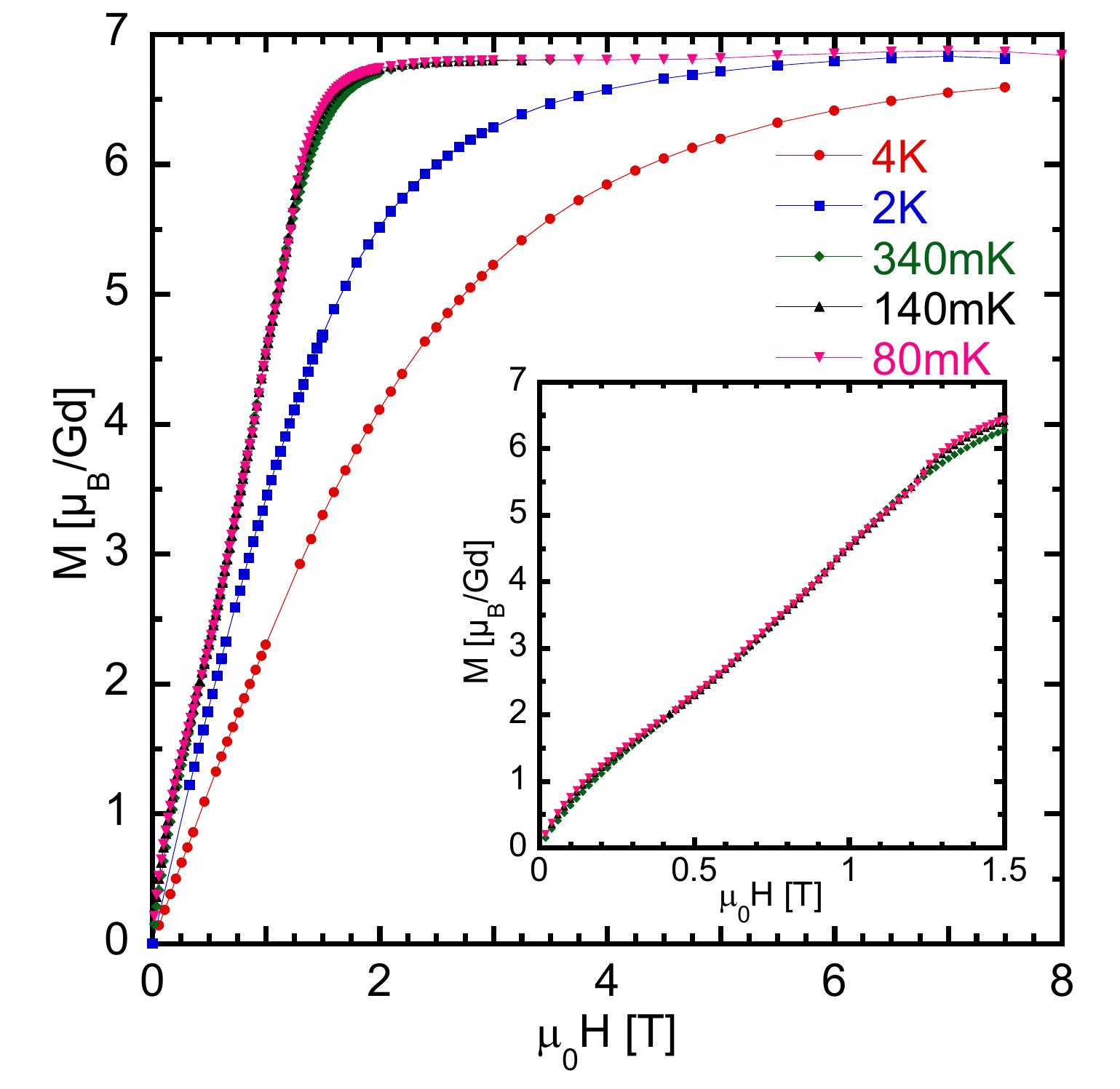}
\caption{$M$ vs $\mu_{0}H$ for 0 $<$ $\mu_{0}$H $<$ 7.5 T with H $\parallel$ to  [1 1 0] for several temperatures between 0.08 and 4.2~K. Inset: zoom on the anomaly between 0 and 1.5 T at low temperature.}
\label{figMH}
\end{figure}

Low field magnetization measurements were performed by cooling down the sample from 4.2~K to 0.08~K in an applied field of 10~mT. The susceptibility $\chi$  ($=M/H$ in low fields for which the magnetization is linear in field) as a function of temperature is shown in Fig. \ref{figXT}. The inset of Fig. \ref{figXT} shows a monotonic decay of the inverse susceptibility with decreasing temperature. A linear behaviour is observed down to 1.5~K and assumes a Curie-Weiss law between 1.5 and 4.2~K that provides a Curie constant of 8.13 emu/mol.K,  consistent with the Gd$^{3+}$ spin moment, $S = 7/2$. The obtained Curie-Weiss temperature $\theta$ is -1.97~K in agreement with previous results \cite{Kinney1979,Schiffer1994}. Below 1 K, the plot deviates from the linear behaviour suggesting the development of correlations between the spins. In Fig.\ref{figMH} isothermal magnetization measurements shows saturation around $M_{sat}$ = 7~$\mu$$_B$/Gd. Above 2~K, no anomaly is seen in the magnetization curves.

\begin{figure} [h]
\includegraphics[keepaspectratio=true, width=8cm]{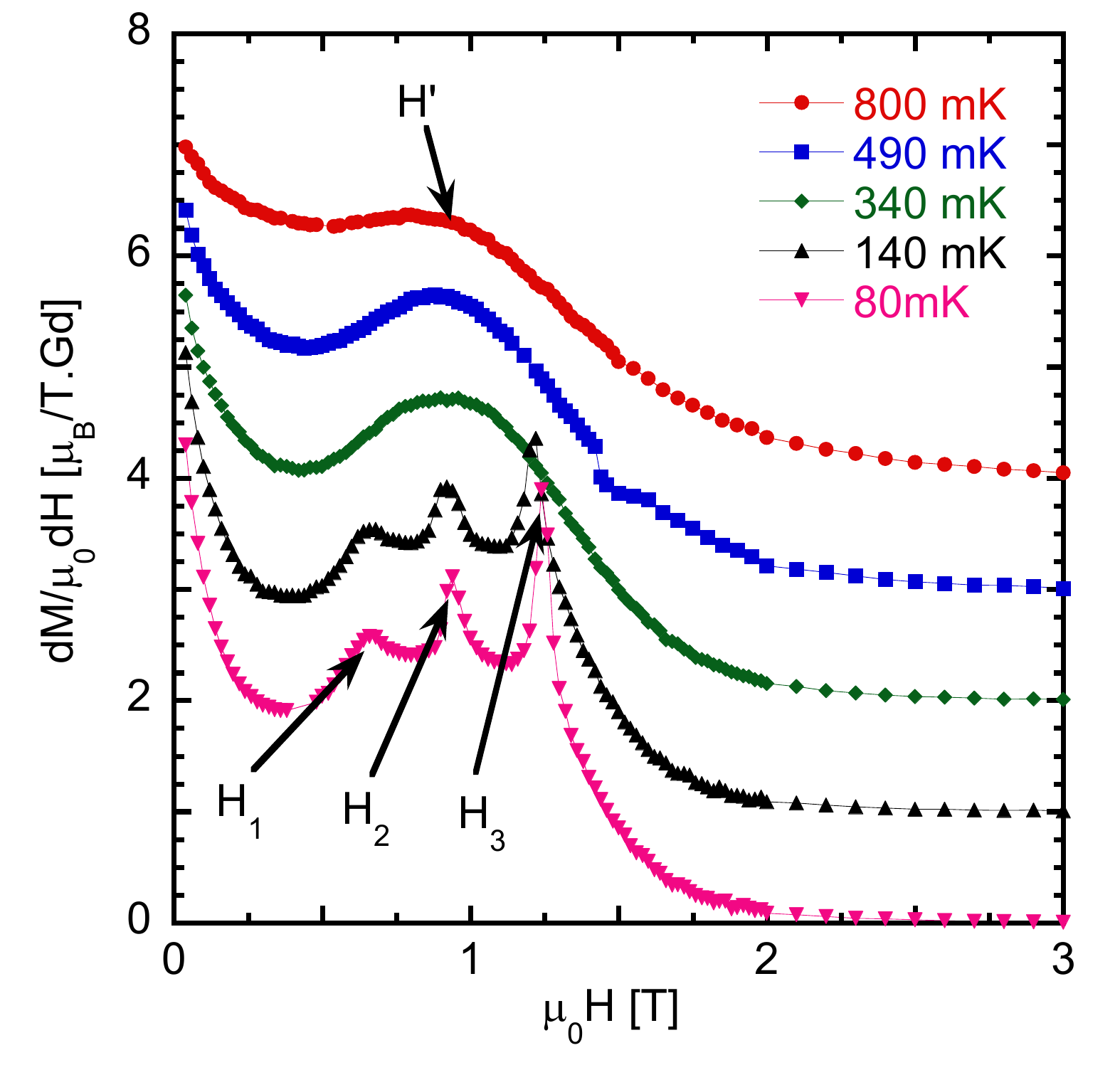}
\caption{$dM/\mu_{0}dH$ vs. $\mu_{0}H$ for a range of applied fields 0~$< \mu_{0}H <$~8~T with H $\parallel$ to [1 1 0]. The curves for different temperatures are separated by 1~$\mu_B$ for clarity. The intensity has been fitted with a Gaussian function for the broad feature and three Lorentzian functions for the three peaks.}
\label{figdMdH}
\end{figure}

A more complex behavior occurs below 1 K. To analyse the $M(H)$ curves as a function of temperature, $dM/dH$ is plotted as a function of the applied field, see Fig. \ref{figdMdH}.
The obtained features show much correspondence with the observations reported by Schiffer  {\it et al.} \cite{Schiffer1994} in heat capacity and susceptibility measurements. 
Fig. \ref{figdMdH} shows the main features with a broad maximum for $H'$ $\sim$ 1 T that develops below 1~K. It has previously been assigned to the quenching of the AF short range order (SRO). There is no significant change in the position or the width of this broad feature with decreasing temperature.  Three well-defined peaks labeled $H_1$, $H_2$ and $H_3$ develop as the temperature is reduced below $\sim$ 0.34~K in addition to the broad feature centered around 1 T. The peaks at the lowest field ($H_{1} = \mu_{0}H$ $\sim$ 0.65~T) and at the highest field ($H_{3} = \mu_{0}H$ $\sim$ 1.25 T) correspond to the boundaries of the previously observed field induced AF long range order \cite{Schiffer1994}. The temperature dependence of the peak positions are in good agreement with previous measurements, but their positions are slightly different possibly the result of anisotropy.  
$H_{2}$ is a very clear and previously  unreported peak at around $H_{2} = \mu_{0}H$ $\sim 0.9$~T. A weak feature was also observed in this field range by Schiffer {\it et al.} \cite{Schiffer1994}, but it was much more rounded and of smaller amplitude.These three peaks,  $H_{1}$,   $H_{2}$ and $H_{3}$, grow on top of the broad SRO feature at $H^{'}$ but do not suppress it, confirming the coexistence between long-range and short-range ordering. It is worth noting that the newly reported $H_{2}$ peak is very sharp, similar to the  $H_{1}$ and $H_{3}$ peaks which were identified as phase boundaries of the field induced ordered phase \cite{Schiffer1994, Hov1980} thereby indicating that $H_{2}$ is the signature of a phase boundary. 
 With increasing temperature the positions of the three peaks and of the broad feature converge towards a single point around $T^* \sim 0.35$~K at the magnetic field $\mu_{0}H^{*} \sim 0.9$~T. To get a deeper understanding of this feature, we performed magnetization measurements as a function of temperature at a constant field.

% To account for the $dM/mu_{0}dH$ behavior at low temperature, it is necessary to assume a Gaussian curve for the broad feature at $H'$ and on top three Lorentzian curves for the three peaks, $H_{1}$, $H_{2}$ and $H_{3}$ that do not diminish the broad feature. Although very specific line shapes have been used to fit the peaks but these are not attributed to any physical phenomena.  

\begin{figure*} [h]
\includegraphics[angle=0, height=5.5cm]{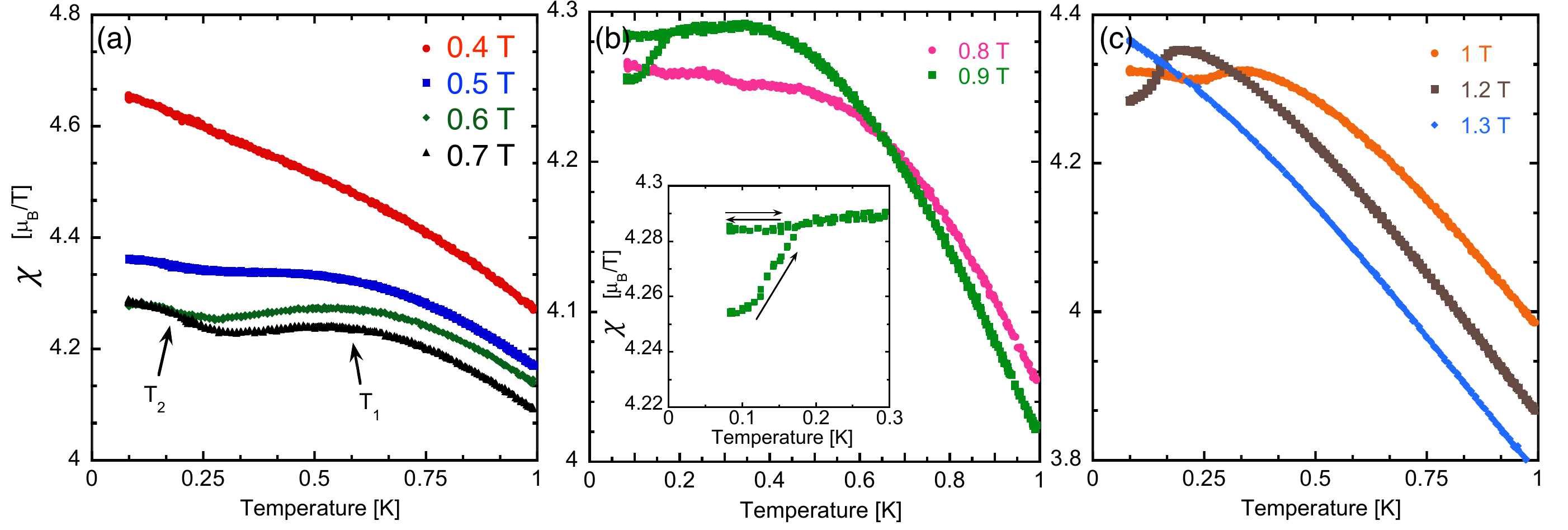}
\caption{$\chi$ vs. $T$ measured with a ZFC-FC procedure for a range of applied fields $0.4 < \mu_{0}H < 1.3$~T along the [1 1 0] direction.}
\label{figMT}
\end{figure*}

At low fields, below 50 mT, the magnetization shows a freezing below 200 mK in the zero field cooled - field cooled (ZFC-FC) measurements, as previously reported \cite{Schiffer1994}. This freezing was also observed in our ac susceptibility measurements. Above 50 mK, this ZFC-FC irreversibility disappears. In the following we focus on larger applied fields to further explore the phase diagram in comparison with our $M$ versus $H$ measurements.   
Fig. \ref{figMT} shows ZFC-FC temperature dependent magnetization measurements for a range of applied magnetic fields, Fig. \ref{figMT}(a):  $0.4 \leq \mu_{0}H \leq 0.7$~T , Fig. \ref{figMT}(b): $0.8 \leq \mu_{0}H \leq 0.9$~T and Fig. \ref{figMT} (c):$1.0 \leq \mu_{0}H \leq 1.3$~T. The magnetic field dependence can be subdivided into five distinct behaviours.
First: $\mu_{0}H < 0.4$~T. ZFC and FC magnetizations overlay and decay monotonically with increasing temperature. No magnetic ordering is observed in this field range.
Second: For $0.5 < \mu_{0}H < 0.7$ T, two distinct features are observed: a broad maximum around $T_{1} = 0.6$~K and a change in slope around $T_{2} = 0.2$ K. $T_{1}$ shifts to lower temperatures when the magnetic field is increased.
Third: For $0.8 < \mu_{0}H < 0.9$ T, the two features overlap so that no clear maximum emerges. A splitting in the ZFC-FC measurements occurs at $\mu_{0}H^{*} = 0.9$ T below $T$ $\sim$ 0.2 K. This ZFC-FC hysteresis is quantitatively reproducible  and is present in a very narrow field region around 0.9~T. The ZFC-FC splitting concerns 7\% of the magnetization.
Fourth: For $0.9 < \mu_{0}H < 1.2$ T,  the curves recovers the shape of the $0.5 < \mu_{0}H < 0.7$~T region.
Fifth: For $\mu_{0}H > 1.3$ T, the magnetization decreases continuously with increasing temperature.

%footnote  0.85 0.95 T there was no variation.
\begin{figure}
 \includegraphics[width=0.54\textwidth]{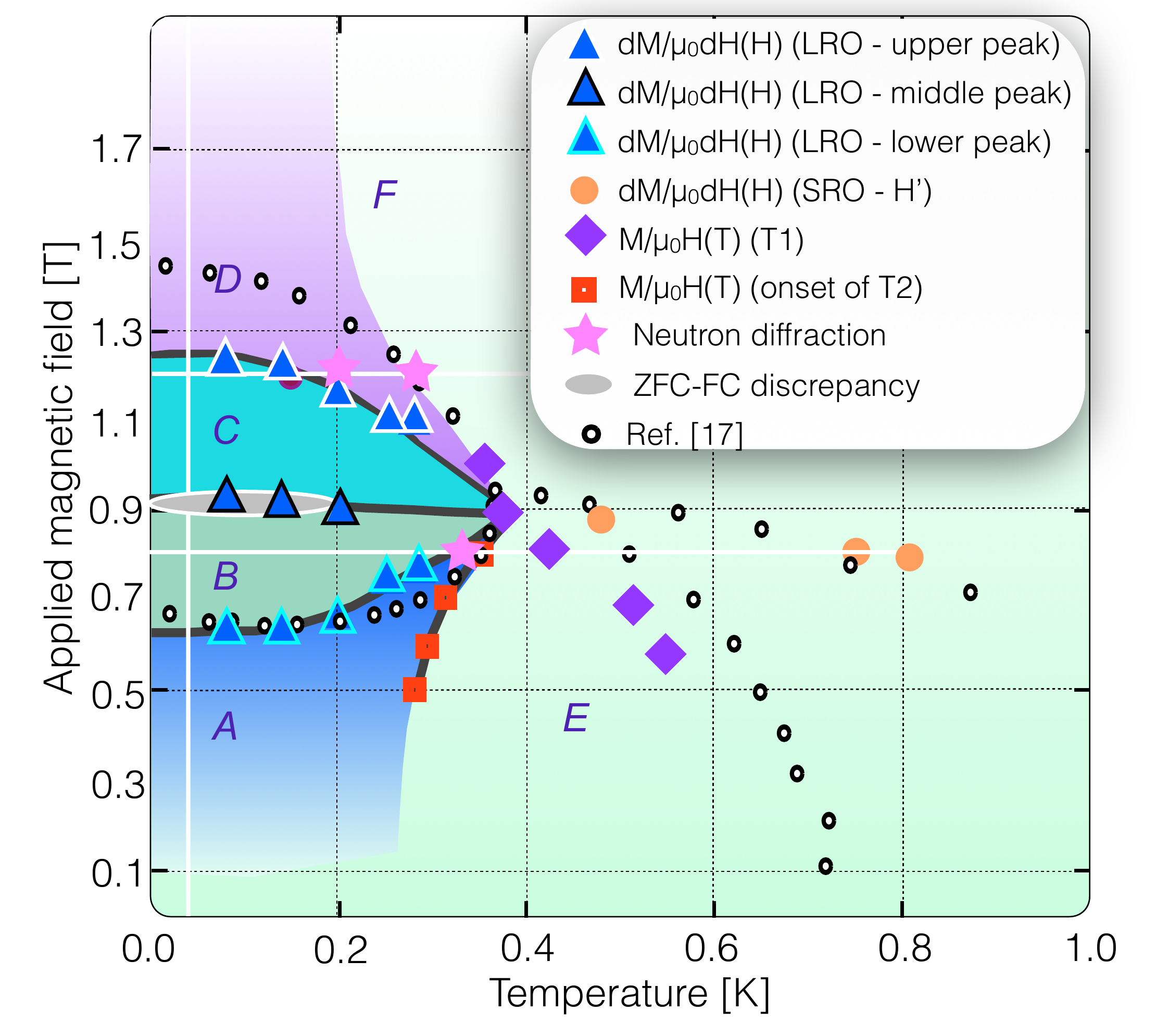}
\caption{The updated phase diagram of Gd$_{3}$Ga$_{5}$O$_{12}$ determined from macroscopic magnetization measurements and neutron powder diffraction. The continuous white lines show regions measured in this study using neutron diffraction.}
\label{PhaseDiagram}
\end{figure}

An updated phase diagram can be constructed from these measurements and is shown in Fig. \ref{PhaseDiagram} and Table \ref{Pdiagram}. High temperatures, $T_{1}$ (orange circles) above $H'$ (violet squares) can be associated with the SRO features previously observed in specific heat and susceptibility measurements by Schiffer {\it et al.} \cite{Schiffer1994}. However, the low temperature change in slope around T$_2$ (red squares) indicates an evolution of the SRO state, which thus can be separated into two regions (A-E) separated at the onset of $T_2$. $H_{1}$ and $H_{3}$ (blue triangles) match the AF field induced region previously observed.  The new observed peak in $M$ vs $\mu_{0}H$ measurements at $H_{2} = \mu_{0}H = 0.9$ T (violet triangles) divides the AF field induced region into two phases (B-C). At this (B-C) phase boundary, $\mu_{0}$H $\sim$ 0.9~T, $T$ $<$ 0.2~K, a small region is reported in which a ZFC-FC discrepancy is observed.  Interestingly this phase boundary  corresponds to the development of field induced phases and the collapse of SRO more explicitly probed using neutron scattering. As such it makes $\mu_{0}H = 0.9$ T, $T \sim 0.35$ K a specific point in the phase diagram of GGG. In the following section these results are compared with the coexistence of short range, incommensurate and AF phases as observed by Petrenko {\it et. al} \cite{Petrenko1998} and extended in this work using neutron diffraction. \\
%%%%%%%%%%%%%%%%%%%%%%%%%%%%%%%%%%%%%%%%%%%%
%%%%%%%%%%%%%%%%%%%%%%%%%%%%%%%%%%%%%%%%%%%%
%%%%%%%%%%%%%%%%%%%%%%%%%%%%%%%%%%%%%%%%%%%%
\begin{center}
\begin{table}
\scriptsize{
\begin{tabular}{ c p{4cm} p{6.5cm} }
\hline
\hline
{\bf Phase} & {\bf M Signature} & {\bf NS Signature}\\
\hspace{1cm} $T$ [K] \hspace{0.5cm}  $\mu_{0}$$H$[T] & & \\
\hline
\hline
{\bf A:} $\big($T $<$ 0.25,  $\mu_{0}H$ $<$ 0.65]$\big)$&   Slope change (T$_{2}$)  &  $\bullet$ Broad feature at 0.8 \AA$^{-1}$ \\
& &  $\bullet$ Broad feature at 0.8 \AA$^{-1}$\\
 & &  $\bullet$ Development of IC and AF peaks \\
 \\
{\bf E:} $\big($[0.25; 1 ], $\mu_{0}H$ $<$ 0.9 $\big)$ &  Broad maximum (T$_{1}$)  & $\star$ Loss of AF (1.14 \AA$^{-1}$) \& IC (1.08 \AA$^{-1}$)\\
& & $\star$ Decrease of SRO \\
\\
{\bf B:} $\big($ T $<$ 0.35, [0.65; 0.9 ] $\big)$ &  Three peaks in $\frac{dM}{\mu_{0}dH}$ &$\bullet$ IC peak $\rightarrow$ [2 1 0] \\
& &$\bullet$ Loss of IC peaks between [1 1 0] $\rightarrow$[$\frac{3}{2}$ 1 $\frac{1}{2}$] \\
 \\
{\bf C:} $\big($ T $<$ 0.35, [0.65; 0.9 ]$\big)$  & Irreversible ZFC-FC&  $\star$ IC peak between [2 0 0] and [2 1 0] fixed in Q\\
& & $\star$ Scattering at Q = 1.69 \AA$^{-1}$ \\
& & $\star$ SRO more correlated\\
\\
{\bf D:} $\big($ T $<$ 0.2, $\mu_{0}$H $>$ 1.2 $\big)$  & No signature &  $\bullet$ Development of scattering at 0.8 \AA$^{-1}$ \\
&  & $\bullet$ Loss of first SRO peak \\
&  & $\bullet$ Appearance of IC peak at Q =  0.80(2) \AA$^{-1}$ \\
& & $\bullet$ Loss of Bragg peak at 1.69 \AA$^{-1}$\\
\\
{\bf F:} $\big($ [0.2;1], $\mu_{0}$H $>$ 0.9 $\big)$  & No signature &  $\star$ Loss of [2 1 0] scattering \\
& & $\star$ Weak variation of [2 0 0] Bragg peak intensity\\
\hline
\hline
\label{Pdiagram}
\end{tabular}
\caption{ Outline of phase diagram presented in Fig. \ref{PhaseDiagram}. SRO = short range order, M = magnetization, NS = neutron scattering.  }
}
\end{table}
\end{center}
%Region D is consistent with the complete loss of the first SRO peak, the appearance of a IC Bragg peak at Q =  0.80(2) \AA$^{-1}$ and the loss of the Bragg peak at Q =~1.69(2) \AA$^{-1}$. 

%\begin{table} [h]
%\begin{tabular}{c c c l}  \hline \hline % *{5}{c}
%\multicolumn{2}{c}{Phase}&Magnetization signature && Neutron scattering signature} \\

%\end{tabular}
%\caption{ Phase diagram }
%\label{Pdiagram}
%\end{table}

%\multicolumn{2}{c}{}&\multirow{1}{*}&AF (1.14{\AA}$^{-1}$) and IC (1.08{\AA}$^{-1}$) disappears \\
%${\bf E:} [0.25;1]$&$H<0.9$&Delimited by a broad maxima (T$_1$) &decrease of magnetic correlations\\
%&&& \\
%\multicolumn{2}{c}{B}&\multirow{1}{*}{ Three clear peaks in \textit{$\frac{dM}{dH}$ vs H }}&\multirow{2}{*}{IC peak moves to AF Bragg position [4 2 0]} \\
%$T<0.35$&$[0.65;0.9]$&& \\
%&&& \\
%\multicolumn{2}{c}{C}&\multirow{1}{*}{Irreversibility in ZFC-FC}&IC peak between [4 0 0] and [4 2 0] is fixed in position \\
%$T<0.35$&$[0.9;1.2]$&&magnetic scattering appears at 1.69{\AA}$^{-1}$ \\
%&&& \\
%\multicolumn{2}{c}{D}&\multirow{1}{*}{No signature}&complete loss of the broad feature at 0.8{\AA}$^{-1}$ \\
%$T<0.2$&$H>1.2$&&AF peak at 1.14{\AA}$^{-1}$ persists for \textit{T}$<$0.3K \\

%&&& \\
%\multicolumn{2}{c
%}{F}&\multirow{2}{*}{no signature}&AF peak at 1.01{\AA}$^{-1}$ persists for $T<0.4K$ \\
%$[0.2;1]$&$H>0.9$&&Bragg peak at 1.69{\AA}$^{-1}$ disappears \\
%\hline \hline
%\end{tabular}
%\caption{ Phase diagram }

%\end{table}

%%%%%%%%%%%%%%%%%%%%%%%%%%%%%%%%%%%%%%%%%%%%
%%%%%%%%%%%%%%%%%%%%%%%%%%%%%%%%%%%%%%%%%%%%
%%%%%%%%%%%%%%%%%%%%%%%%%%%%%%%%%%%%%%%%%%%%

\section{Neutron diffraction}

This study focusses on the regions of the phase diagram encompassing $0.175 < T < 3$~K ($\mu_{0}H = 0$ T) and $0.06 < T < 3$~K ( $0 < \mu_{0}H < 2$ ~T). The unusual scattering observed for $T < 0.14$~K in zero field with longer ranged order superposed on short range correlations is not considered. Neutron scattering corroborates the complicated phase diagram shown in Fig. \ref{PhaseDiagram} with strong consistency between  magnetization and neutron diffraction measurements. \\

Petrenko {\it et al.} \cite{PetrenkoHFM} concluded that competing magnetic interactions result in competition between various ground states and prevents GGG from ordering in zero magnetic field while applying a small magnetic field crystallises the magnetic state into ordered components with a range of ferromagnetic (FM), AF and incommensurate (IC) propagation vectors. 

%In contrast muon spin relaxation and M\"{o}ssbauer spectroscopy indicate a temperature spin relaxation  indicative of slow spin fluctuations down to 25~mK and up to 1.8~T \cite{Dunsiger2000, Marshall2002, Bonnville2004}. The contradiction between a long range ordered and dynamic magnetic state can be understood in terms of the time scales probed by the various techniques such that the system appears paramagnetic on the fast timescale of the muon but looks static within the energy resolution of the neutron. This has been observed in the frustrated magnet Tb$_{2}$Ti$_{2}$O$_{7}$ for which it was shown that there were many length and timescales to consider \cite{TTO2012}.

The present work provides a greater insight into the interplay between different magnetic orders. Particular emphasis is placed on the perturbation of the short range order with respect to the other magnetically ordered states. Short range magnetic order pervades the $(H, T)$ phase in which uniquely long range magnetic correlations have previously been assigned.  The interplay between  short range, FM, AF and IC orders are correlated with the phase diagram outlined in Fig. \ref{PhaseDiagram} and with the main results condensed in Table \ref{Pdiagram}. These datasets highlight why a refinement of the magnetic structure with a single wavevector has eluded previous studies.
%We moved this part. Not sure whether it is a good idea. 

\subsection{Experimental details} 

Neutron scattering experiments were performed on the D7 diffuse scattering spectrometer and the cold chopper spectrometer IN5 at the Institut Laue Langevin (ILL), Grenoble. The sample has previous been used in the work of Petrenko  {\it et al.} \cite{Petrenko1998} and Deen {\it et al.} \cite{Deen2010} containing 99.98\% of the non-absorbing isotope $^{160}$Gd. The sample was covered with Iso-Propanol (or 2-Propanol) 99\% Deuterium that freezes the crystallites into place without any substantial contribution to the scattering.

The zero field data, measured on IN5, shows the scattering within the elastic resolution of the instrument,  80 $\mu$eV full width at half maximum (FWHM). The energy resolution is determined using a standard incoherent scatterer, see section \ref{sectionFieldDependence}. The IN5 data set could not be integrated in energy due to  spurious scattering in parts of the inelastic spectrum.

The field dependent data has been measured using D7 with E$_{I}$ =  3.55 meV. The wide angular range available on D7 enables simultaneous determination of short and long ranged scattering and elucidate their interplay. Field dependent measurements up to 2.5~T were performed at 0.05~K with a temperature dependence  measured, up to 0.8~K, at  0.8 and 1.2~T, as depicted by the white continuous lines on Fig.\ref{PhaseDiagram}. Uniaxial polarization analysis performed under applied field on D7 provides two scattering cross sections,   spin-flip and non-spin flip scattering. The spin-flip
scattering  is entirely magnetic in origin since the spin incoherent cross-section for GGG is negligible.  The non spin-flip scattering contains nuclear and
magnetic components. A separation of the spin-flip from the non-spin flip contributions allows the nuclear scattering to be isolated while the spin-flip contributions provide the magnetic cross-sections \cite{StewartD7}. The data have been corrected for detector and polarization  analyzer efficiencies using standard samples of vanadium and amorphous silica respectively \cite{StewartD7}. All data are corrected for background contributions by subtracting the scattering from an empty sample can under equivalent conditions. The D7 data set is integrated in energy. Nevertheless the data sets from D7 and IN5 can be compared since previous work show  that the majority of the scattering is elastic (82\%) and the features of interest for this study lie solely in the elastic line \cite{Deen2010, Deen2013}. This is discussed further in the appendix.

\subsection{Ambient field behaviour}

\begin{figure}
\includegraphics[width =\textwidth]{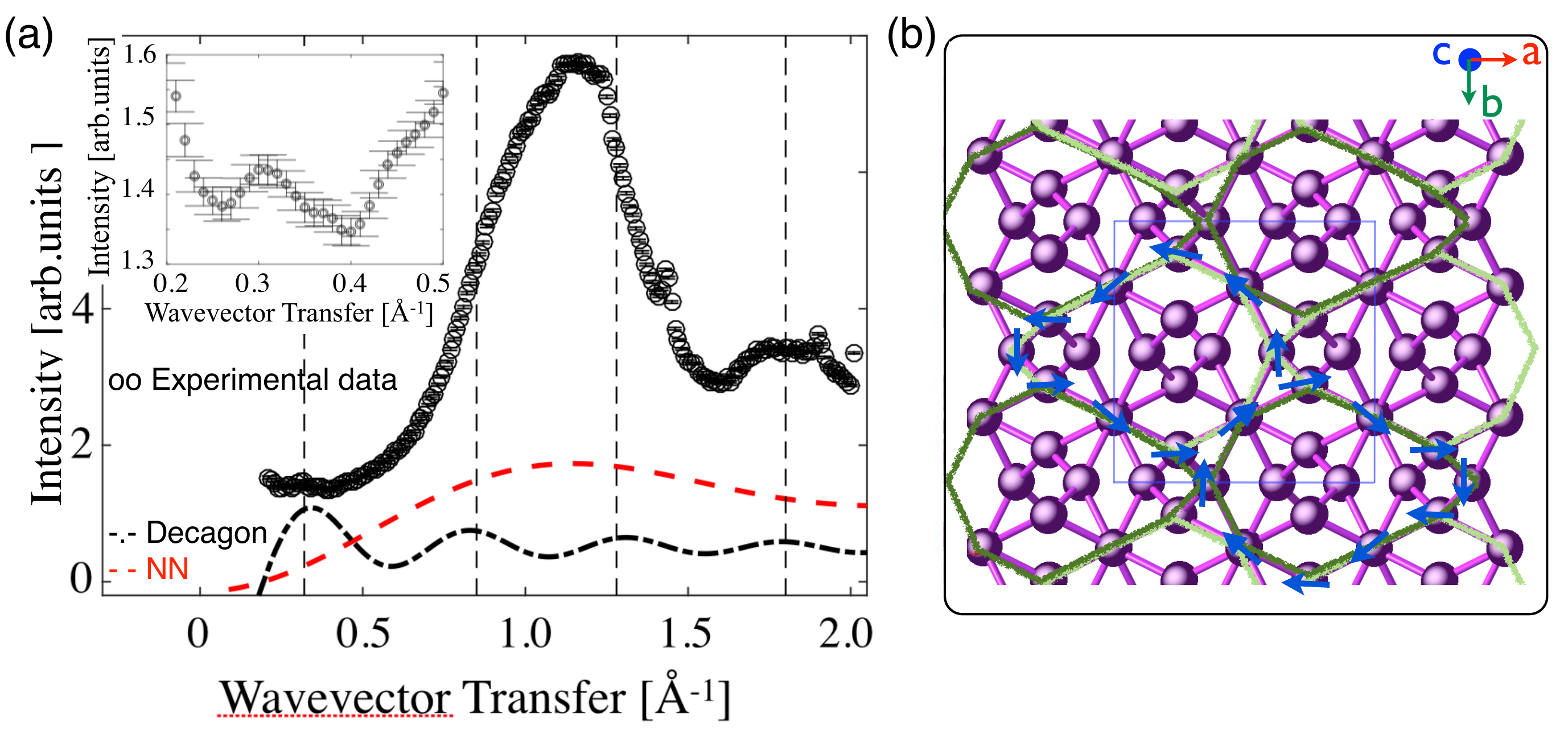}
\caption{(a) Neutron diffraction of Gd$_{3}$Ga$_{5}$O$_{12}$ at 0.25 K, 0 T with a near neighbor (NN) spin model (red line --) and exchange between ten ion spin clusters (black -.-) for comparison. Inset: Zoom of the magnetic scattering around Q = 0.3 $\AA^{-1}$. (b) Snapshot of ions involved in the ten ion spin cluster for the [1 0 0]/[0 1 0] crystallographic  direction. The blue rectangle is the structural unit cell. The light green and light blue loops show two domains. The arrows denote the relative spin directions of the ions.}
\label{TenIonDataModel}
\end{figure}

Region A, depicted in the phase diagram of Fig. \ref{PhaseDiagram},  has generally been considered to consist of short range magnetic correlations. Petrenko {\it et al.} \cite{PetrenkoRMC} determined, using Monte Carlo (MC) methods, near neighbor interactions, J$_{1}$ = 0.107~K,  as the dominant magnetic exchange term.  Upon closer inspection of the scattering profiles, see Fig. \ref{TenIonDataModel}(a),  it can be seen that a model with only near-neighbor exchange cannot give rise to the scattering profile observed. Instead,  it is proposed that a self organised super-lattice unit, a ten ion  spin cluster, is required to account for some of the features, see Fig. \ref{TenIonDataModel}(b). The arguments behind such a model are presented in the next few paragraphs.\\
%that obey Heisenberg near neighbour exchange on a triangle, namely $\sum_{triangle} S$ = 0 with near neighbor spins at 120$^{o}$ relative to each other
 Short range magnetic correlations in a powder sample scatter according to the expression
\begin{equation}
\left({\frac{d \sigma}{d \Omega}}\right)_\text{Mag} \sim  \sum_{n} \frac{\left<{\bf S_{0}}\cdot {\bf S_{n}}\right>}{S(S+1)}N_{n}\frac{\sin(QR_{n})}{QR_{n}}F(Q),
\label{SRO}
 \end{equation}
where $S_{0}$ and $S_{n}$ are the spin magnitudes  of the central atom and the $n^{\rm th}$ shell atom, $R_{n}$ and $N_{n}$ are the radii and coordination numbers of the $n^{\rm th}$ nearest neighbor shell respectively  \cite{Rainford} and $F(Q)$ is the magnetic formfactor \cite{BrownTables, Squires}.  Fig. \ref{TenIonDataModel}(a) shows the expected magnetic scattering profile for short range order, using eqn. \ref{SRO}, in  comparison to the magnetic scattering profile measured at $T = 0.25$~K (0~T). Interestingly, the short range order profile does not reproduce all the features expected, in particular, the broad feature centered on Q = 1.8 \AA$^{-1}$.  In order to reproduce scattering at Q = 1.8 \AA{} magnetic exchange  between ten ion spin clusters is considered. 

In recent years there has been much excitement surrounding emergent behaviour in geometrically complex materials. The kagome lattice has provided such an emergent structure in the form of correlated hexagon loops of spins \cite{ZnCr2O4Nature2004Lee}. It has been suggested that the first and second order zero energy modes on the hyperkagome structure are dynamic loops of spins that include triangular loops of spins, trimers, and ten ion spin clusters, decagons\cite{Bergholtz2010, Zhitomirsky2008, Robert2008}. The interpretation from recent optical hole-burning experiments supports the existence of ten ion spin clusters in the low temperature spin glass phase of GGG \cite{Ghosh2008}. Fig. \ref{TenIonDataModel}(b) shows a snapshot, in time, of possible spin directions involved in such a ten ion spin cluster for the [1 0 0]/[0 1 0]  crystallographic directions. The loops are decoupled from their environments and suggested to be  quantum protectorates \cite{LaughlinPines, Kuzemsky}. The magnetic exchange for trimers can be modelled using eqn. \ref{SRO} with only near neighbor exchange and the requirement that Heisenberg spins on a triangular lattice  $\sum_{triangle} S$ = 0 resulting in near neighbour spins 120$^{o}$ relative to each other.  The magnetic neutron scattering profile for a trimer, see Fig. \ref{TenIonDataModel}(a),  shows a maximum in scattering intensity at the same Q position as near neighbor magnetic correlations, Q  $\sim$ 1.1 \AA$^{-1}$. The spins on a decagon loop also obey the Heisenberg near neighbor spin requirement yet the spins are correlated throughout the ten ion spin cluster.  The exchange between ten spin ion clusters can be described using eqn. \ref{SRO} to consider short range interactions between adjacent clusters. The coordination number of a ten ion spin cluster is 14 with an average unit cell size of 13.12~\AA. The magnetic scattering from a ten ion spin clusters, see Fig. \ref{TenIonDataModel}(b), provides many extra features in common with the  powder diffraction data of GGG at 0.25~K and 0~T.  \\
 % Try to clarify this paragraph. Explicitely say that trimers give the same peak as NN. (maybe should be rephrased). 
 % figure in In5July2010/BasicClusterRings.m 
The models show that using a combination of near neighbour and exchange between ten spin ion clusters can explain some experimental features. First, around Q $\sim$ 0.3~\AA$^{-1}$ there is a weak feature consistent with the position of the first peak of the ten ion spin cluster, see inset Fig. \ref{TenIonDataModel}(a). Second, the main peak of the near neighbor model is found at the same position as the main peak in the data, however this peak is strongly asymmetric and this can be ascribed to scattering from ten ion spin clusters which provides two features either side of the main peak position, see dashed lines in Fig. \ref{TenIonDataModel}(a). Thirdly, a strong scattering feature at 1.8~\AA$^{-1}$ in the ten ion spin cluster model is consistent with the data. Interestingly, the magnetic correlation length $\xi = 2\pi/ \Delta Q$ of the SRO peak at 1.8 \AA$^{-1}$ corresponds to an average unit cell size of the ten ion spin clusters, $\xi$ = 13.8$\pm$2~\AA{}. 
A combination of these two models captures the position of the scattering if not perfectly the respective weights  of the scattering. Nevertheless it provides a strong indication of unusual correlations such as the coexistence of near neighbor exchange with a ten ion spin cluster. This dataset therefore captures the first two soft modes of spins on hyperkagome structure, dynamic trimer and decagon spin loops as a feature of region A. %The magnetic scattering correlation length of the first SRO peak is calculated via $\xi = 2\pi/\Delta Q$. Interestingly $\xi$ at 0.05~K and 0.8~T is 13.09(3) \AA{}, comparable to the average dimension of the ten ion cluster. 
% figure in In5July2010/BasicClusterRings.m

\newpage
\subsection{Magnetic behaviour under applied magnetic fields. }
\subsubsection{Field dependence (T = 0.05~K)}
\label{sectionFieldDependence}
\begin{figure}[htp]
\includegraphics[width = 0.45\textwidth]{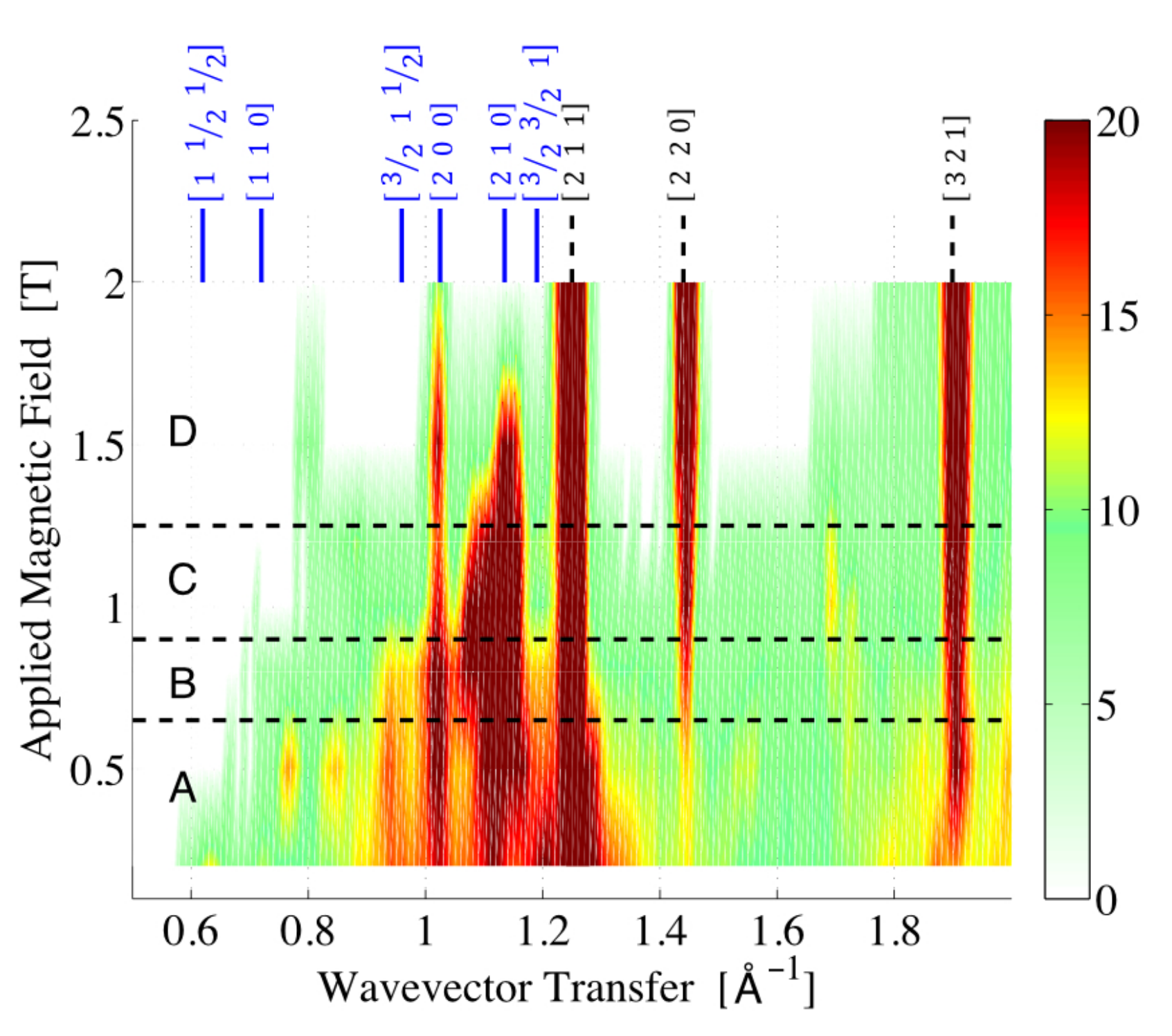}
\caption{Overview of the field dependence of the magnetic neutron scattering at 0.05~K. Peaks corresponding to AF peak positions are marked with full vertical blue lines while FM peak positions are marked by dashed vertical black lines. The phase boundaries found in Fig. \ref{PhaseDiagram} are presented by dashed horizontal lines.}
\label{FieldDependence2D}
\end{figure}

An overview of the field dependence of the magnetic scattering is shown in Fig. \ref{FieldDependence2D} and reveals great complexity. Peaks corresponding to AF peak positions are marked with continuous blue lines while FM  peak positions are marked by dashed black lines. The phase boundaries reported on Fig. \ref{PhaseDiagram} are reproduced on Fig. \ref{FieldDependence2D}. Variations in the magnetic scattering profiles can be accurately mapped onto the phase diagram as described in Table \ref{Pdiagram}. A closer look at the individual scattering profiles for each applied magnetic field provides more detailed information.\\

%/Users/pascaledeen/Deen/Deen_ILL_2006_2010/deen/Gd3Ga5O12/D7Data/D7_2008/FieldDep.m

\begin{figure}[htp]
\includegraphics[width = 0.45\textwidth]{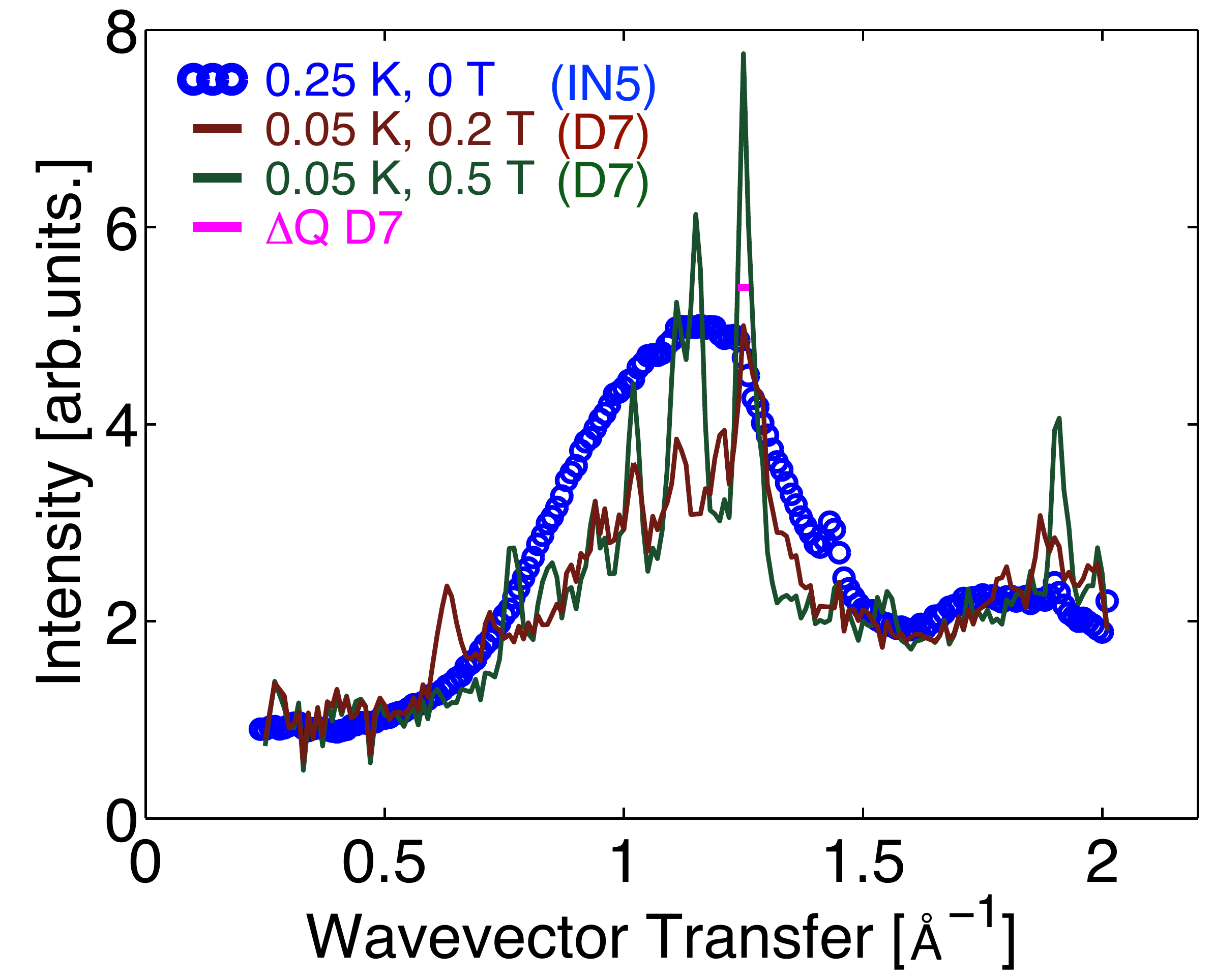}
\caption{(a)Field dependence of magnetic neutron diffraction at 0.25~K \& 0~T, 0.05~K \& 0.2~T and 0.05 K \& 0.5~T.}
\label{FieldDependence}
\end{figure}

Fig. \ref{FieldDependence}(a) shows magnetic scattering profiles in region A $\rightarrow$ H for 0, 0.2 and 0.5~T ($T < 0.25$~K). The scattering profile at 0.25~K,  0~T is the elastic scattering within the energy resolution, 80 $\mu$eV of IN5, further information is provided in the appendix.Upon the application of only 0.2~T the short ranged ordered peaks are drastically reduced. This is particularly true for the first SRO peak at  Q $\sim$ 1.1~\AA$^{-1}$. Interestingly, the second SRO peak, assigned to scattering from a decagon loop in the previous section, is more robust and remains unaffected up to 0.5~T. This would indicate that decagons are more resilient to perturbation than the trimer counterparts upon the application of a magnetic field.  Longer ranged correlations are developed by 0.2~T and are resolution limited at 0.5~T, see Fig. \ref{FieldDependence}.

%REMOVED The Q-resolution of D7, measured using a calibration sample, Y$_{3}$Fe$_{5}$O$_{12}$, is shown for comparison

%The base temperature (T = 0.05~K) field dependence of the various scattering contributions reveal strong interplay between SRO, AF and IC contributions. 
%PlotdataFebruary2013.m
%Magnetoelasticcoupling_ D7_2008
\begin{figure}[htp]
\includegraphics[width = \textwidth]{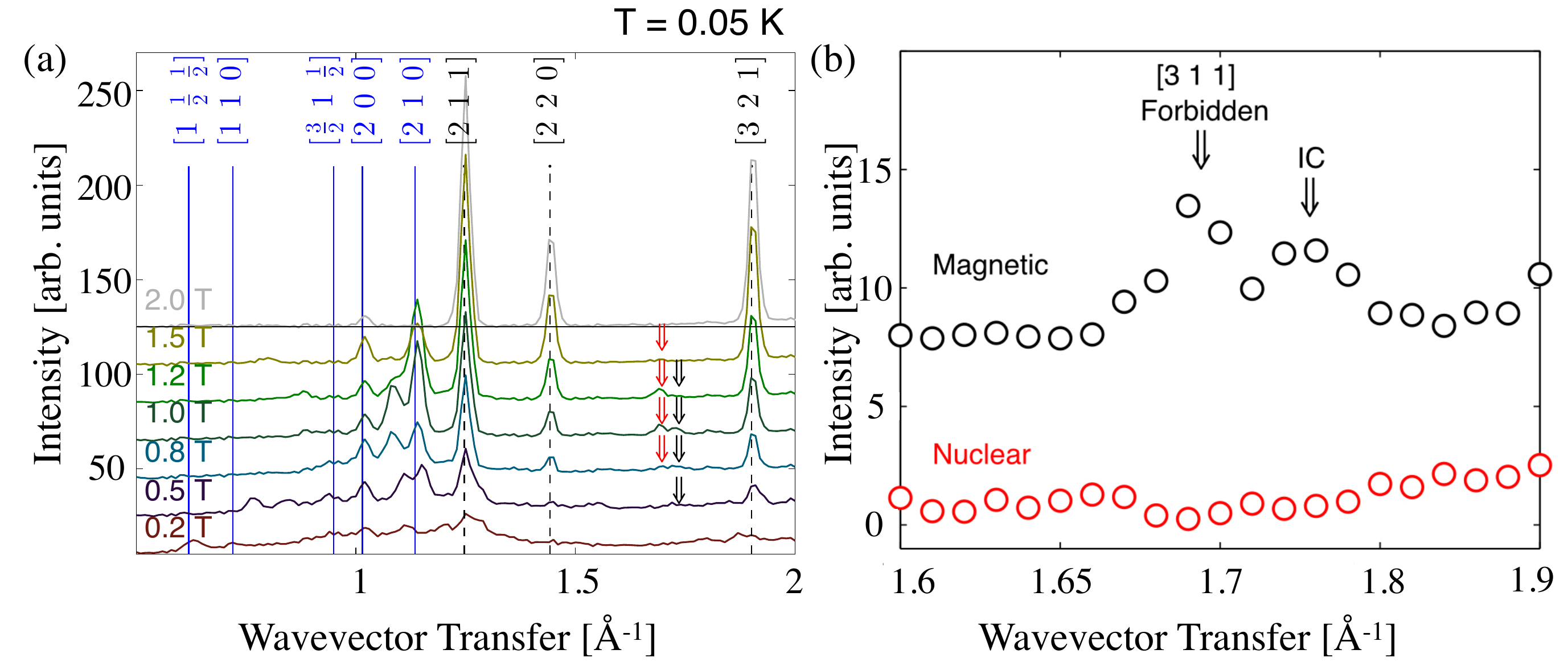}
\caption{(a) Field dependence of the magnetic scattering at 0.05~K. The continuous blue vertical lines correspond to order that is commensurate with the crystal structure and could represent AF order. Dashed black vertical lines are nuclear/ FM positions. The vertical arrows denote scattering at the forbidden nuclear scattering position [3 1 1] and an IC position. The data are offset for clarity. (b) Scattering profiles of the magnetic and nuclear contributions at 0.05~K and 1~T zoomed in on the region of the scattering at 1.69 and 1.72~\AA$^{-1}$. The nuclear profile shows no equivalent scattering.}
\label{BaseTempFieldDependence}
\end{figure}

Fig. \ref{BaseTempFieldDependence}(a) shows the complete field dependence of the magnetic diffraction patterns at base temperature, 0.05~K.  FM correlations are not considered in this work since it is impossible to disentangle the effect of local magnetic exchange interactions and the increase in scattering at the nuclear Bragg peak positions due to the application of a magnetic field on a powder. The data are offset for clarity. The continuous vertical lines correspond to positions in reciprocal space where AF order is expected while dashed vertical lines are positions corresponding to nuclear/FM order.  In region A, 0 $<$ $\mu_{0}$H $<$ 0.65~T, the IC features observed between the [1 1 0] and [$\frac{3}{2}$ 1 $\frac{1}{2}$] Bragg peak positions and for
Q $<$ 1.2~\AA$^{-1}$, are very fluid with respect to their absolute position in reciprocal space. This is particularly true for the broad features around 0.6 - 0.8 $\AA^{-1}$ at 0.2~T that have shifted to between 0.7 - 0.9 $\AA^{-1}$ by 0.5~T and are not observed beyond.  The boundary between phases A and B fixes the Bragg positions of various AF and IC Bragg peaks. An example is the [2 1 0] Bragg peak position which appears to move from an IC position in phase A to the [2 1 0] Bragg peak position in phase B. Additionally, the IC scattering between [2 0 0] and [2 1 0] is also fixed into position as the A-B phase boundary in crossed. Upon the establishment of phase C, 0.9 $<$ $\mu_{0}$H $<$ 1.2~T, magnetic scattering for Q~$<$~0.8~\AA$^{-1}$ is diminished and magnetic scattering appears at  Q =~1.69(2) $\AA^{-1}$.  Interestingly this scattering corresponds to the [3 1 1] Bragg peak position, forbidden in neutron scattering for this crystallographic symmetry. This scattering could then be the result of either a magnetic phase change or a signature of magneto elastic coupling (MEC). MEC can be a very weak effect, possibly  the result of a slight rotation of oxygen around the dodecahedral Gd$^{3+}$ site.  However, MEC should also give rise to scattering in the nuclear channel however neutron scattering might not be sufficiently sensitive to reveal the nuclear scattering part. Indeed there is no nuclear scattering peak  at the [3 1 1] Bragg peak position when comparing the magnetic scattering cross section to the nuclear scattering cross section, see  Fig. \ref{BaseTempFieldDependence}(b), so the origin of this scattering remains unclear. \\

\begin{figure}[htp]
\includegraphics[width = 0.45\textwidth]{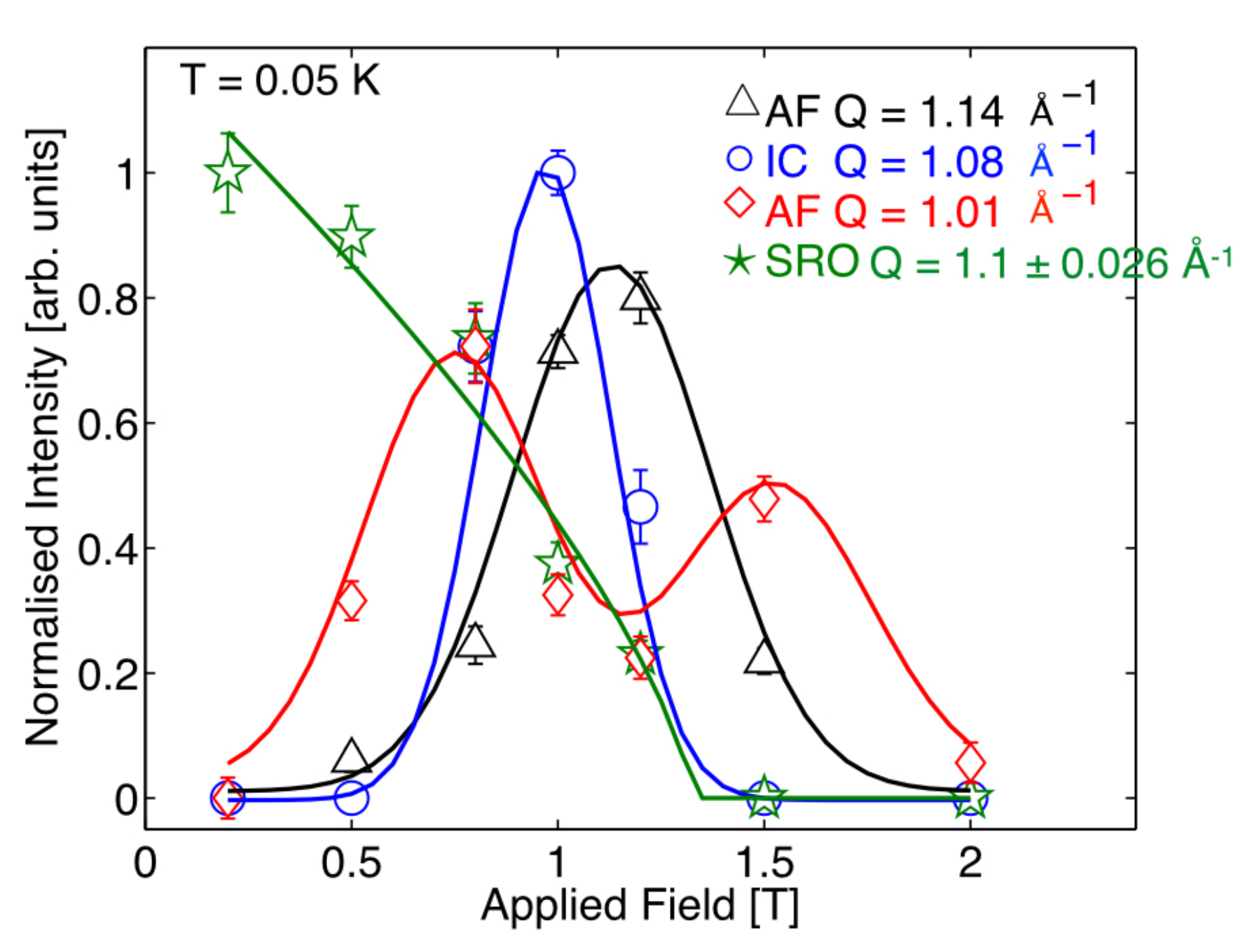}
\caption{Field dependent integrated intensities of AF, IC and short range order. The data is normalised for ease of viewing and the lines through the data are to guide the eye.}
\label{FieldDependenceB}
\end{figure}

Region D is consistent with the complete loss of the first SRO peak, the appearance of a IC Bragg peak at Q =  0.80(2) \AA$^{-1}$ and the loss of the Bragg peak at Q =~1.69(2) \AA$^{-1}$. Although the first SRO peak has disappeared in this region, the second SRO peak, at Q = 1.8 \AA$^{-1}$, remains strong up to and beyond 2.0~T. Above 1.5 T, all AF peaks are suppressed except the AF peak at 1.01~\AA{}  ([2 0 0]) which persists. 
% main changes in this paragraph: add some precisions on the Q value. Suppress some statements which appeared two times. Suppress reference to the D* phase which is not anymore in the phase diagram. 
\\
An overview of the field dependent behaviour of the integrated intensities of some of the magnetic scattering is given in Fig. \ref{FieldDependenceB}. It is particularly interesting to note the interplay between the various magnetic contributions such that  the AF scattering at Q = 1.14~\AA$^{-1}$, ([2 1 0]), is most intense when the AF scattering at  Q = 1.01~\AA$^{-1}$  ([2 0 0]) is much reduced only for the Q = 1.14~\AA$^{-1}$ peak to restrengthen when the Q = 1.01~\AA$^{-1}$ peak diminishes, region D. In addition, the field dependence of the IC peak at 1.08~\AA$^{-1}$ follows the integrated intensities of the AF peak at Q = 1.14~\AA$^{-1}$. 
 %PlotDataForPhase.m

% Figures to be found in plotdataFebruary2013B, plotdataFebruary2013_more,PlotDataforPhase.m

\subsubsection{Temperature dependence at 0.8~T (T = 0.05~K)}

\begin{figure}[htp]
\includegraphics[width = \textwidth]{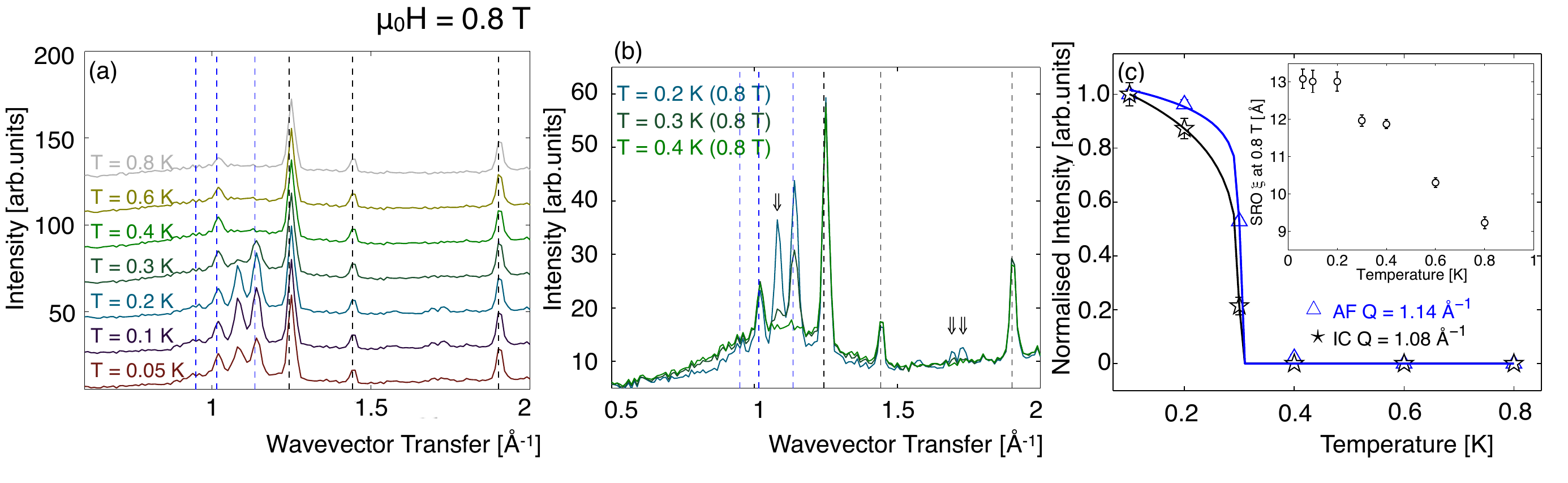}
\caption{ (a) Temperature dependence of the magnetic scattering at 0.8~T. The data are offset for clarity. The blue dashed vertical lines corresponding to AF order. Dashed vertical lines are nuclear/ F positions. (b) Focus on the region 0.2$< T <$ 0.4~K (0.8~T). The black arrows emphasise the greatest variations in this temperature range. (c) Temperature dependence of the integrated intensities of the AF (Q =  1.14  \AA$^{-1}$ ) and IC (Q = 1.08 \AA$^{-1}$ )   clearly showing the phase transition A $\rightarrow$ E defined in  Fig. \ref{PhaseDiagram}. Inset: Temperature dependence of $\xi$ for the first SRO peak.}
\label{TempDep0_08T}
\end{figure}

The magnetization data implies a convergence phases to a unique point around 0.9~T and 0.35~K. Temperature dependent  powder diffraction has been measured at 0.8~T to determine the unique nature of this region, see Fig. \ref{TempDep0_08T}(a,b), with particular emphasis on the region $0.2 < T < 0.4$~K. These data  show a very distinct transition at 0.3~K with both AF and IC orders disappearing beyond 0.3~K, see Fig. \ref{TempDep0_08T}(c). The line separating regions A $\rightarrow$ E is observed in the neutron scattering profiles. The magnetic scattering correlation of the first SRO peak broadens beyond 0.2~K from 13.1(3) \AA{} at the lowest temperatures to 9.2(2) \AA{} at 0.8~K, see inset of Fig. \ref{TempDep0_08T}(c). It is difficult to extract $\xi$ for the second short range ordered peak due to its weaker nature and limited Q-range.  The magnetic scattering at the IC position, Q = 1.73 \AA$^{-1}$, and the forbidden nuclear Bragg peak position Q = 1.69 \AA$^{-1}$  disappear between  0.2 and 0.3~K, see the black arrows on Fig. \ref{TempDep0_08T}(b). These data do indeed indicate that there is a convergence of phases around 0.9~T and 0.35~K as indicated in Fig. \ref{PhaseDiagram}, however more detailed studies are needed to elucidate the precise details of this region.

%width at 0.8 T 0.05 K  1.4056e-01    9.7734e-03
%0.4 K 1.8182e-01    3.4267e-02

\subsubsection{Temperature dependence at 1.2~T}

\begin{figure}
\includegraphics[width = \textwidth]{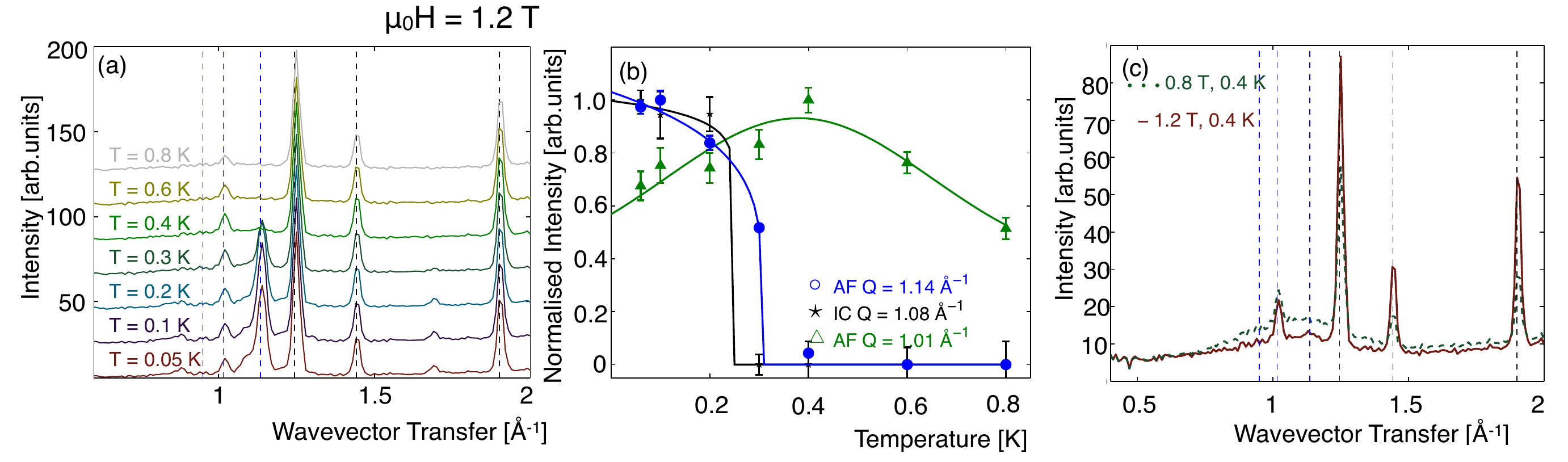}
\caption{(a)Temperature dependence of Gd$_{3}$Ga$_{5}$O$_{12}$ at 1.2~T. The data are offset for clarity. The blue dashed  vertical lines corresponding to AF order. Black dashed vertical lines are nuclear/ FM positions. (b) Integrated intensity of AF and IC Bragg peaks. Lines are a guide to the eye. The arrows show regions are greatest variation in this temperature range. (c) Neutron scattering profiles across the phase boundary E $\rightarrow$ F. No clear transition is apparent. }
\label{TempDep1_2T}
\end{figure}

The neutron scattering data of the 1.2~T temperature dependence, Fig. \ref{TempDep1_2T}, traverses regions C $\rightarrow$ D $\rightarrow$ F.    The IC Peak at 1.08 \AA~disappears between $0.2 < T < 0.3$~K ,  black arrow in Fig. \ref{TempDep1_2T}(b), region C $\rightarrow$ D. The peak at the position of the forbidden nuclear Bragg peak, Q  =  1.69 \AA$^{-1}$, in addition to the AF Bragg peak at Q = 1.14 \AA$^{-1}$ disappear between $0.3 < T < 0.4$~K, region D $\rightarrow$ F. Interestingly the peak at Q = 1.01 \AA$^{-1}$, a peak that can be attributed to AF order, is only slightly perturbed in the $0.3 < T < 0.4$~K region. Fig. \ref{TempDep1_2T}(c) shows the variation in scattering profiles across the F $\rightarrow$ E transition. A slight variation in intensity is observed but no clear transition can be observed via neutron diffraction. Single crystal measurements are required to elucidate the details of the variation in scattering to determine whether the regions E and F are distinct phases. 
%The distinct change of phase into phase F is not captured by magnetisation data.
% suppress this sentence because we do see something between E and F in magnetisation, but add a sentence above concerning the D-F boundary. 

\section{Conclusion}
An investigation of the applied magnetic field and temperature dependent behaviour of the archetypal magnetically frustrated compound Gd$_{3}$Ga$_{5}$O$_{12}$ is presented. Single crystal and powder magnetic susceptibility in conjunction with polarised neutron diffraction reveal a unified picture of  the phase diagram with close correlation between magnetic susceptibility measurements with the field applied along the [1 1 0] crystalline direction and neutron scattering profiles of a powdered sample. \\

Several extra phase boundaries are required to correctly describe the (H,T) phase diagram of GGG with H $\parallel$ [1 1 0] . There is strong evidence that trimerised and decagon loops of spins coexist at low temperatures and low magnetic fields. These emergent loops of spins are strongly affected by the long range ordered components yet coexist with incommensurate and AF order up to 1.3~T for spin trimers and beyond 2.0~T for decagon spin structures. Magnetization measurements allude to a multiphase convergence around 0.9~T and 0.35~K. Interestingly, there is a strong ZFC-FC discrepancy at 0.9~T and $T$ $<$ 0.2~K that is very well defined in applied field indicative of a further glassy phase. The close link between susceptibility and neutron powder scattering thus indicates that the behaviours observed exist throughout the crystal, originates from the spin Hamiltonian on the hyperkagome lattice and are an intrinsic feature of frustration on hyperkagome structures.

%Figure \ref{TenIonDataModel}(b) In particular it is surprising that SRO remains so prominent up to 1.5~T and seems to be most closely related to the behaviour of the IC order.

\section{Acknowledgements}
The project was partly funded by the Danish Re-449
search Council for Nature and Universe through DANSCATT. P.D. would like to thank the sample environment group at the ILL for their support and H. Mutka and O.A Petrenko for stimulating discussions.  We would like to thank C. Paulsen for the use of his magnetometers and J. Debray for the orientation of the single crystal.
O. F. acknowledges a grant from the Laboratoire d'excellence LANEF in Grenoble.

\bibliographystyle{unsrt}
\bibliography{GGG_2306.bib}

\end{document}